\documentclass[sigconf]{acmart}

\acmConference[AdKDD 2022]{}{August 15,
  2022}{Washington, D.C.}
\acmDOI{}
\acmPrice{}
\acmISBN{}

\usepackage{amsmath,amssymb,amsfonts}
\usepackage{multirow}
\usepackage{algorithm, algorithmic}
\renewcommand{\algorithmicrequire}{\textbf{Input:}}

\newcommand{\argmax}{ \mathop{\rm arg~max } \limits }

\usepackage{graphicx}
\usepackage{xcolor}

\def\BibTeX{{\rm B\kern-.05em{\sc i\kern-.025em b}\kern-.08em
    T\kern-.1667em\lower.7ex\hbox{E}\kern-.125emX}}

\author{Shion Ishikawa}
\email{shion.ishikawa@rakuten.com}
\affiliation{%
  \institution{Rakuten Institute of Technology}
  \country{Japan}
}

\author{Young-joo Chung}
\email{youngjoo.chung@rakuten.com}
\affiliation{%
  \institution{Rakuten Institute of Technology}
  \country{The United States of America}
}

\author{Yu Hirate}
\email{yu.hirate@rakuten.com}

\affiliation{%
  \institution{Rakuten Institute of Technology}
  \country{Japan}
}

\begin{abstract}
Recently online advertisers utilize Recommender systems (RSs) for display advertising to improve users’ engagement. Contextual bandit model is a widely used RS to exploit and explore users’ engagement and maximize the long-term rewards such as clicks or conversions. However, the current models aim to optimize a set of ads only in a specific domain and do not share information with other models in multiple domains. In this paper, we propose dynamic collaborative filtering Thompson Sampling (DCTS), the novel yet simple model to transfer knowledge among multiple bandit models. DCTS exploits similarities between users and between ads to estimate a prior distribution of Thompson sampling. Such similarities are obtained based on contextual features of users and ads. Similarities enable models in a domain that didn’t have much data to converge more quickly by transferring knowledge. Moreover, DCTS incorporates temporal dynamics of users to track the user’s recent change of preference. We first show transferring knowledge and incorporating temporal dynamics improve the performance of the baseline models on a synthetic dataset. Then we conduct an empirical analysis on a real-world dataset and the result showed that DCTS improves click-through rate by 9.7\% than the state-of-the-art models. We also analyze hyper-parameters that adjust temporal dynamics and similarities and show the best parameter which maximizes CTR.
\end{abstract}

\begin{document}

\title{Dynamic collaborative filtering Thompson Sampling for cross-domain advertisements recommendation}


\maketitle

\section{Introduction}

Recently online advertisers utilize Recommender systems (RSs) for display advertising to improve users’ engagement. Traditionally, collaborative filtering \cite{su2009survey} is one of the popular recommender systems \cite{anastasakos2009collaborative}. However, it suffers from the cold start problem \cite{silva2019pure}; Collaborative filtering cannot make a good recommendation for new items and new users without historical interactions. Content-based methods and hybrid models \cite{DBLP:conf/recsys/Kula15},\cite{guo2017deepfm} leverage contextual features (i.e., side information) of items and users to overcome the cold-start problem.

Meanwhile, reinforcement learning (RL) -based methods have shown its effectiveness on Recommender systems. By utilizing both exploration and exploitation, it can maximize the long-term profits for RS. It is suited for advertisement recommendation, where the ads are shown in the limited time period. Among reinforcement learning-based methods, bandit models were frequently applied to that problem \cite{pandey2006handling}, which dynamically selects arms (corresponding to actions in RL) and maximizes total rewards. Contextual bandit that utilizes side-information of users and ads as contexts (e.g., user demography, creative characteristics) achieved great success in various areas such as movie and book recommendation. (LinUCB \cite{li2010contextual} and LinTS \cite{agrawal2013thompson}).
Ads, unlike movies and products, are created when new sales campaigns begin and are removed from displays when campaigns are over. Therefore, ratios of newly created ads are higher compared to new movies or products. Traditional methods such as collaborative filtering can't perform well for ad-recommendation because of the sparsity of training data and it is not suited for long-term profit optimization. RL solved long-term profit problem and can deal with cold-start cases relatively well, but it still suffers from this extreme cold-start conditions. In this scenario, the agent tends to focus on exploration and sacrifice short-term reward for long-term reward. The $\epsilon$-first policy \cite{tran2010epsilon}, for instance, purely explores in cold-start conditions via random recommendations, which goes overboard with the exploitation and exploration balancing and results in even worse performance. Here previous models were mainly created to optimize a specific widget on a web domain and each model was independent of each other. Considering that the number of active users in a domain (a business unit) varies in large companies with multiple business units, it is natural to transfer model’s knowledge from popular domain to unpopular/new domains to improve the recommendation performance and convergence speed. Thus, in this paper we propose a model to improve performance of existing RL-based recommend systems by utilizing knowledge transfer.

\begin{figure}
  \centering
  \includegraphics[width=7.5cm]{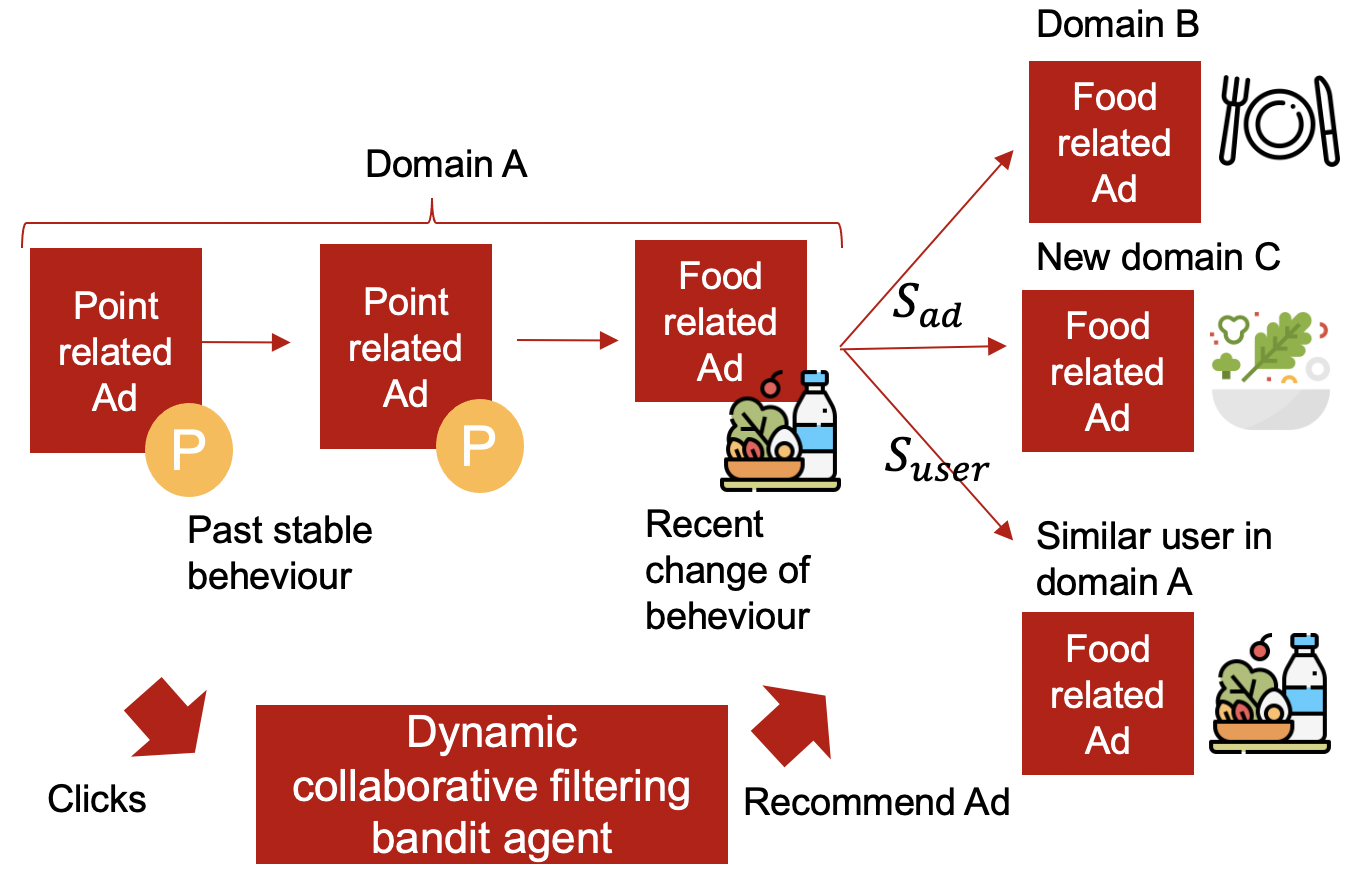}
  \caption{Use-case of the dynamic collaborative filtering bandit.}
  \label{fig:introduction}
\end{figure}

We therefore propose multi-domain optimization to allow all model to communicate and share knowledge with each other. Our method, dynamic collaborative filtering Thompson Sampling (DCTS), can accelerate convergence with only a small amount of data by utilizing information from models in different domains. We here naturally expanded the concept of similarities among ads \cite{slivkins2011contextual} \cite{li2016collaborative}  to those in different domains. In addition, we added the concept of temporal dynamics by incorporating temporal dynamics to reflect user's sudden preference change. Recently collaborative filtering with dynamic terms \cite{koren2009collaborative} and a RNN-based model \cite{ma2020temporal} have improved existing performances. We here added the concept of temporal dynamics by incorporating discounting rewards \cite{kocsis2006discounted}.

 In summary, our proposed model incorporates temporal dynamics in the multi-domains and it enables our model to track the entire user's journey among multi-domains. We showed one use-case of multi-domains optimization in Fig. \ref{fig:introduction}. For example, a user was interested in earning reward points and clicked point-related advertisements. Recently this user came to get an interest in eating delicious food and clicked food-related ads on the e-commerce site (domain A). Our model will recommend the related restaurant in the restaurant reservation website (domain B), can recommend related hotels in the newly created website for lodging reservations (domain C), and improve the cold start issue of domain C. Our model also recommends food-related ads clicked by the user to the similar user in domain A. These kinds of cross-domain and cross-user recommendations will be conducted based on similarity function $S_{ad}$ and $S_{user}$. To evaluate the performance of our model for this kind of use-case, in this research, we first compare the performance of our model with the one of existing bandit models on synthetic dataset, and show our model leveraged knowledge in other domains and adapt the sudden change of customer's preference. Then, we empirically showed our model could also outperformed existing bandits on the real-world dataset. Here we used our internal dataset of online ads in our Rakuten Ichiba and Rakuten Travel , and observed improvement of CTR.
 
 Our contributions are as follows:
\begin{enumerate}
      \item We propose a dynamic collaborative filtering Thompson Sampling (DCTS), which improve the estimation of prior distribution for Thompson Sampling by utilizing similarity among ads and users, and discounting rewards. DCTS leverages contextual features of ads and users for similarity computation. Our model will leverage this similarity and transfer knowledge across heterogeneous domains. It leads to track recent crosss-domain behavior of users and gives a prior estimation of arms. As a result, it accelerates convergence.
      \item We conducted empirical analyses on real-world dataset for advertisements in multiple domains. We compared our method with the state-of-the-art algorithms such as Transferrable LinUCB \cite{labille2021transferable}, hybrid LinUCB, and Thompson Sampling. As a result, DCTS performed better than T-LinUCB by 9.7 \% and better than LinUCB and TS by around 37 \% in terms of click-through rate (CTR).
      \item We also evaluated the performance of hyper-parameters. Regarding the parameter $\gamma$ which adjusts decays of rewards, we observed the best performance when $\gamma$ = 0.25. It was better than when using responses in the latest batch ($\gamma$ = 0) and no decay ($\gamma$ = 1). Regarding the parameter $\gamma$ which adjusts the importance of global rewards, the best values when $g$ =1 or $g$ =0.1. We assume this is because real-world dataset are sparse. This suggests it's still important to retain global rewards even when personalized rewards are considered.

\end{enumerate}
 
\section{Related Work}
In this section, we discuss the existing bandit models for recommend systems. First of all, Thompson sampling \cite{chapelle2011empirical} was introduced to select an arm that maximizes the expected reward in a randomly drawn belief. It picked the best arm based on the sampled score from the posterior distribution. Upper-Confidence-Bound methods deterministically take the action according to an optimistic estimateof the reward they will yield. LinUCB \cite{li2010contextual} proposed the contextual bandit models to improve the rewards. 

To incorporate temporal dynamics with bandit models, UCB with discounting reward was proposed \cite{kocsis2006discounted}. It utilized discount to reflect recent change of user's behaviour. Then, Rotting Bandit \cite{levine2017rotting} was proposed. Rotting Bandit handles discount as a function of the number of times each arm has been pulled. These previous studies showed that incorporating dynamic change of user's preference could increase the accuracy for the Yahoo! new recommendation and KDD Cup 2012 Online Ads. 

 To add pre-estimation of rewards by utilizing similarities, contextual zooming algorithms \cite{slivkins2011contextual} and bandit with collaborative filtering \cite{li2016collaborative} were proposed. Bandit with collaborative filtering utilized similarities between items in one domain, also dynamically built clusters of users and items and leveraged them for the calculation of similarities. Our research expanded the concept of bandits with similarities to the dynamic optimizations across multiple heterogeneous domains. Also, we utilized Locally-sensitive Hashing \cite{charikar2002similarity} to calculate similarities efficiently.
 
As for bandits with transfer learning, TCB \cite{liu2018transferable} introduced the translation matrix among domains to utilize similarities of domains. T-LinUCB \cite{liu2018transferable} utilized this similarities as prior estimation of domains and alleviated cold-start problems.
Here we describe the difference between our model and TCB/T-LinUCB. First, our model is an expansion of Thompson sampling whereas TCB/T-LinUCB were expansions of UCB; Second, our model leverages similarity between users as well as domains; Finally, only our model incorporates dynamic behaviour, captures recent change of preference in the cross-domains. Also, B-kl-UCB \cite{zhang2017transfer} is a bandit considering transfer learning but this models assume the set of actions are same across domain; on the other hand, our model can handle different action set in the heterogeneous domains by leveraging similarities.

\section{Model}
In this section, we propose the dynamic collaborative filtering Thompson sampling (DCTS). First, we assume we have $N$ available sources. Each source is corresponded to a widget where ads are displayed. In each source, we have a set of ads. Let a set of ads in source $s$ be $\mathbf{A_s}$. Also let $\mathbf{X}$ be a matrix of $m$ users which have $d_u$ features where $\mathbf{X} \in \mathbb{R}^{m \times d_u}$, and let $\mathbf{Y^s}$ be a matrix of $k_s$ ads which are in source $s$ and have $d_a$ features where $\mathbf{Y^s} \in \mathbb{R}^{k_s \times d_a}$. Then, we denote $i_t^s$ as user $i$ in source $s$ at time step $t$. For each time step $t$ and source $s$, we observe $x_{i_t^s} \in \mathbf{X}$ and $y_{a^s_t} \in \mathbf{Y^s}$ as contexts, and user will see ad $a^s_t \in \mathbf{A_s}$ and we observe reward $r_{i_t^sa_t^st}$ where $r_{i_t^sa_t^st}$ is an implicit reward which indicates whether user $i_t^s$ clicked ad $a^s_t \in \mathbf{A_s}$ or not. We here denote observation as $O_t^s = (x_{i_t^s}, y_{a^s_t}, r_{i_t^sa_t^st})$.

The objective of DCTS is to pick up an ad $a_t^s$ to display as an action to maximize cumulative rewards $\sum_s^N\sum_{t=0}^{T} r_{i_t^sa_t^st}$ where T is a maximum time step. This can be represented as minimizing total regret of user $i$ as follows:
\begin{equation}
\begin{split}
{\text{minimize }} \mathbb{E}[\text{regret}_i(T)] = 
\sum_{s=0}^N\mathbb{E}[\underset{a_t^s \in A_s}{\text{max}}\sum_{t=0}^T r^*_{i_t^sa_t^st} - \sum_{t=0}^T r_{i_t^sa_t^st}]
\end{split}
\end{equation}
where $r^*$ indicates a reward from the best action for user $i_t^s$.
\subsection{transferring knowledge via similarities of sources and users}
DCTS learns a policy to pick up ads from observations of all ad sources $\mathbf{O} = \{O^s\}_{s=0,...,N}$. DCTS utilizes connections between sources and transfers knowledge among sources and users, which leads a policy to be aware of more generalized user's behaviors.
DCTS transfers rewards among users and among ads based on their similarities. Here we utilize cosine similarity as a degree of transferring:
\begin{equation}
\begin{split}
\label{eq:similarity}
\mathcal{S}_{user}(x_i, x_j) = \frac{x_i \cdot x_j}{|x_i||x_j|}
\end{split}
\end{equation}
 where $x_i$ and $x_j$ are contextual features of user $i$ and $j$. In the same way, we define $\mathcal{S}_{ad}(y_i, y_j)$ by contextual features of ads.

The number of users is huge in the real-world dataset and calculating similarities among all pairs of users is not scalable. Therefore, we leveraged Locally-sensitive Hashing \cite{charikar2002similarity} to obtain similarities of users efficiently.
\subsection{Dynamic collaborative filtering Thompson sampling}

Before describing an algorithm of DCTS, we first describe the original Thompson sampling.
Thompson sampling in the bandit problem is a method for pulling an arm that maximizes the expected reward. This is equal to sampling estimated scores from the posterior distributions of the arm in each round and choosing the arm with the highest score. In Bernoulli Bandit case, the likelihood function is formulated by Bernoulli distribution whose prior distribution is Beta distribution as a natural conjugate prior. Beta distribution can be written as
\begin{equation}
\begin{split}
\mathcal{B}(\theta_k; \alpha_k, \beta_k) = \frac{\mathbf{\Gamma}(\alpha_k + \beta_k)}{\mathbf{\Gamma}(\alpha_k)\mathbf{\Gamma}(\beta_k)} \theta_k^{\alpha_k-1}(1-\theta_k)^{\beta_k-1}
\end{split}
\end{equation}
where $\theta_k$ is a probability that action $a_k$ produces rewards, and $\alpha_k$ and $\beta_k$ are parameters. As a non-informative prior, previous studies assumed the special case where $\alpha_k = 1$ and $\beta_k = 1$. \emph{However, for the most practical cases, values of actions can be estimated by utilizing historical data.} With leveraging user-user and ad-ad similarity, DCTS offers the better estimation of prior distribution.
First, we formulate the pre-estimated parameters for each user $i$ and ad $k$ at time step $t$ as follows:

\begin{equation}
\begin{split}
\alpha^0_{ik}(t) = \sum_{l\neq k} S_{ad}(y_k, y_l)s_{il}(t) + \sum_{j \neq i} S_{user}(x_i, x_j)s_{jk}(t)
\\
\beta^0_{ik}(t) = \sum_{l \neq k} S_{ad}(y_k, y_l)f_{il}(t) +  \sum_{j \neq i} S_{user}(x_i, x_j)f_{jk}(t)
\label{eq:preparams}
\end{split}
\end{equation}
where $s_{il}(t)$ is a discount-aware cumulative reward
\begin{equation}
\begin{split}
s_{il}(t) =  \sum_{\tau=0}^{t} \gamma^{t-\tau} s_{ij\tau}
\end{split}
\end{equation}
and $s_{ij\tau}$ is a binary variable, which will be $1$ if we observed reward for user $i$ and ad $k$ at $\tau$ and $0$ otherwise, and $\gamma$ indicates a discount ratio. In the same way, $f_{jk}(t)$ will be defined as a discount-aware cumulative negative reward by $f_{ij\tau}$ and a discount ratio $\gamma$. $f_{ij\tau}$ represent the number of failure of recommnedation; it will be $1$ if user $i$ viewed ad $k$ at $\tau$ but didn't click the ad. Here we naturally assume the value of ad $k$ for user $i$ can be estimated by similarity between them. Therefore, we transferred rewards based on similarities of users and ads. We also assumed user's preferences will change with time, so we discount rewards and put a high value on rewards from user's recent behaviors.

Second, we formulate posterior distribution. For the basic Thompson sampling, posterior distribution will be formulated by Beta distribution. With uniform prior knowledge, the parameters were $\alpha_{k} = s_k + 1$ and $\beta_{k} = f_k + 1$. In this research, with using prior knowledge in Eq. \ref{eq:preparams}, we formulate parameters of posterior Beta distribution as follows:

\begin{equation}
\begin{split}
\label{eq:params}
\alpha_{ik}(t) = \lambda(s)\alpha^0_{ik}(t) + g s_{k}(t) + s_{ik}(t) + 1
\\
\beta_{ik}(t) = \lambda(f)\beta^0_{ik}(t) + g f_{k}(t) + f_{ik}(t) + 1
\end{split}
\end{equation}
where $\lambda$ is a hyper-parameter which adjusts importance of prior knowledge and $\lambda(s) = \frac{\lambda}{s_{ik}(t)+1}$, $\lambda(f) = \frac{\lambda}{f_{ik}(t)+1}$.
$g$ is a hyper-parameter that adjusts the importance of global rewards. Like the original Thompson sampling, we introduce $s_k(t)$ and $f_k(t)$ as a mean of rewards among users, since the interactions of users and ads are sparse so that there are few valid similar users will occur in real-world cases. 1 at the last term is a pseudo count to avoid errors when historical reward were unavailable. We present details of the policy DCTS in Algorithm 1.

 \begin{algorithm}
 \caption{Dynamic collaborative filtering Thompson sampling}
 \begin{algorithmic}[1]
 \renewcommand{\algorithmicrequire}{\textbf{Input:} $\lambda, g, \gamma \in \mathbb{R}^0_+, S_{user}, S_{ad}$, Source observations $\mathbf{O}$}
 \REQUIRE
  \FOR {$t=0,...,T$}
  \STATE{Observe user $i_t^s$ and context $x_{i_t^s}$, action sets $\bf{A_s}$ and their contexts $\bf{Y^s}$}
  \FOR {$k \in \bf{A_s}$}
  \STATE {Calculate $\alpha^0_{i_t^sk}(t)$ and $\beta^0_{i_t^sk}(t)$ according to Eq. \ref{eq:preparams}}

  \STATE{Calculate $\alpha_{i_t^sk}(t)$ and $\beta_{i_t^sk}(t)$ according to Eq. \ref{eq:params}}
  \STATE{Sample $\theta_k$ from the $\mathcal{B}(\theta_k; \alpha_{i_t^sk}(t)$, $\beta_{i_t^sk}(t))$}
  \ENDFOR
  \STATE Play action $k = \underset{k}{\argmax}$ $\theta_k$ and observe reward $r_{i_t^skt}$
  \STATE Add observation  $O^s_t \gets (x_{i_t^s}, k, r_{i_t^skt})$
  \ENDFOR
 \end{algorithmic} 
 \end{algorithm}


\section{Experiments}
\subsection{Synthetic Experiment}
  
In this section, we evaluated the two advantages of DCTS with a simple synthetic environment. The first advantage is a performance of transferring knowledge among multi-domains. We prepared 10 ads and displayed 5 of the ads on domain A, and the remaining 5 ads on domain B. In the first 500 steps, there is only domain A and users were only allowed to visit domain A. In the latter 500 steps, domain B was created and users were allowed to visit domain B. We evaluated the performance when users first visited domain B. This kind of scenario often happens in the real world when users newly registered domain B, or when domain B or ads in domain B are newly created. In this synthetic setting, we set $\lambda=1$, $g=1$, and $\gamma=1$. As for $\gamma$, we wanted to investigate the performance of transferring, so we disabled discounting rewards and set $\gamma$ as $1$. We also prepared the user's response function as an environment of reinforcement learning. This function receives contextual attributes of users and ads and returns rewards with the probability, which is obtained from the logistic regression with contextual attributes. Weights of logistic regression were sampled from the standard normal distribution. 

\begin{figure}
  \centering
  \includegraphics[width=7cm]{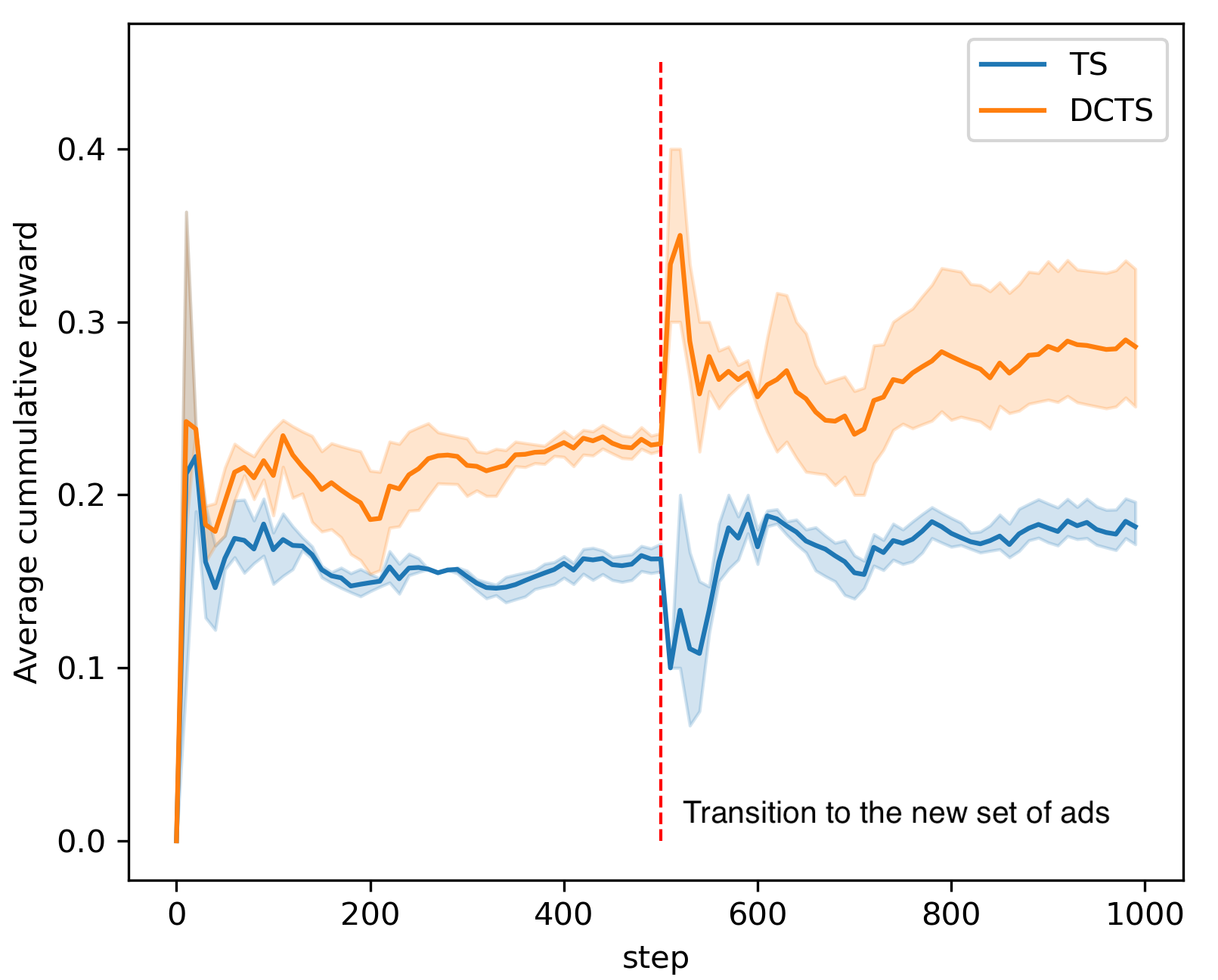}
  \caption{Synthetic simulation of transferring knowledge in 2 different pages}
  \small{We drew a line plot of the average cumulative reward. The red vertical line indicates the time step we changed a set of ads. The shaded band indicates 95\% confidence interval obtained by three times.}
  \label{fig:synthe_transfer}
\end{figure}

\begin{figure}
  \centering
  \includegraphics[width=6.5cm]{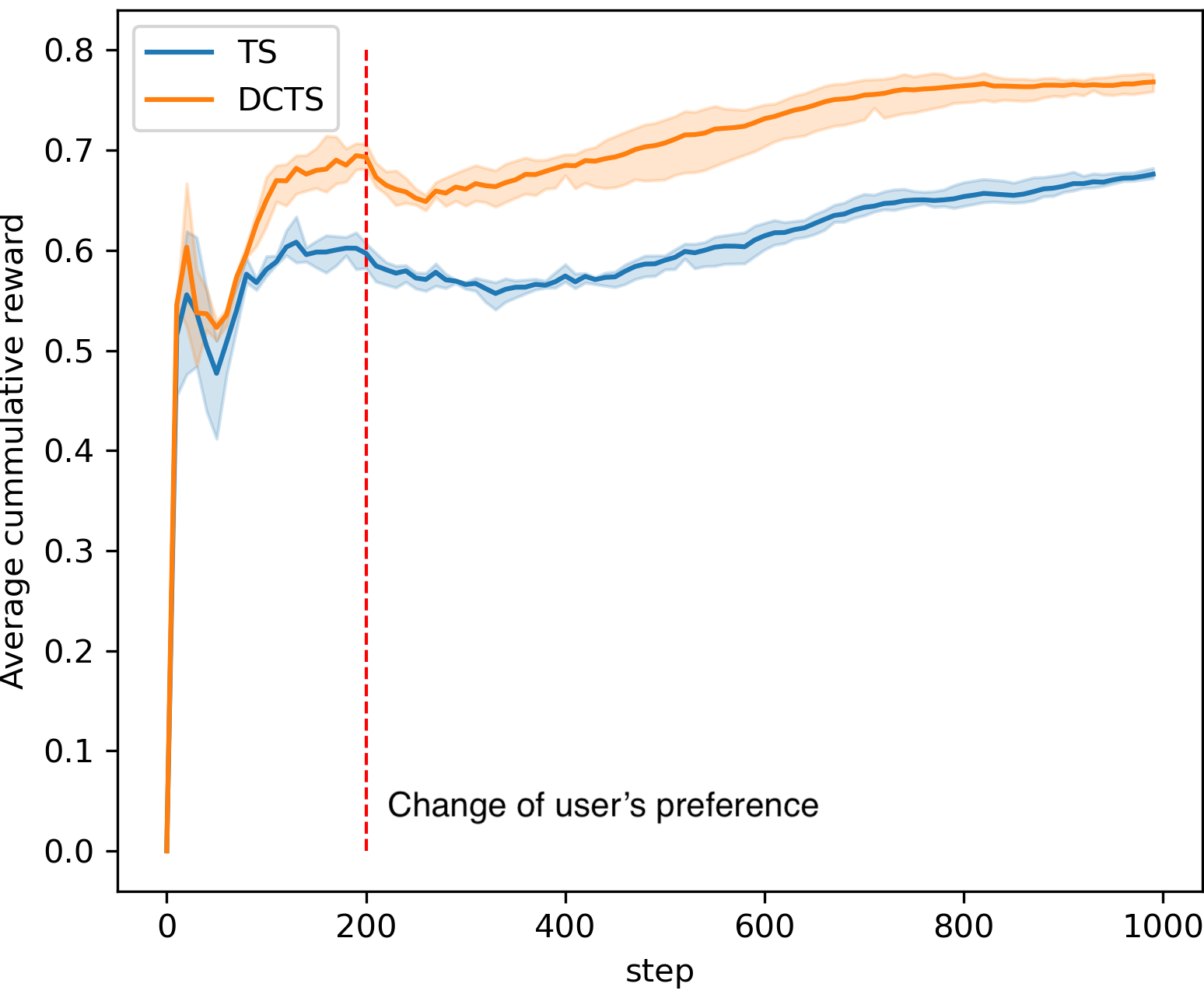}
  \caption{Synthetic simulation of changing user's preference}
  \small{ We drew a line plot of the average cumulative reward across time steps. The red vertical line indicates the time step when the change of user's preference occurred. The shaded band indicates 95\% confidence interval obtained by three times executions.}
  \label{fig:synthe_dynamic}
  \end{figure}
According to Fig. \ref{fig:synthe_transfer}, at the very beginning TS and DCTS had the same performance but over time DCTS outperformed than TS because DCTS could handle a part of contextual preferences as personal rewards $s_{ik}$ and $f_{ik}$. Also, as we assumed, the performance of DCTS right after the transition to the new set of ads was much better than that of TS. This is because DCTS was able to leverage knowledge of previous ad, while TS had to start exploration again.
 We assume the average rewards of DCTS decreased until time step 700 because the weight of prior knowledge $\lambda(s(t))$ decayed. In this case, higher $\lambda$ seems to be better. Later in this paper, we will investigate the effect of hyper-parameters.

To verify the second advantage of DCTS, namely its fast adaptation to user's preference change, we prepared 50 ads and made a user's response function, which is the same as the above one but its contextual weights will change at the specified time. In this synthetic setting, we set $\lambda=1$ and $g=0$ and $\gamma=0.95$. We let the discount of rewards occur at every 10-time step because in the real-world setting we update models in a batch and the time step will be a certain duration of time. According to Fig. \ref{fig:synthe_dynamic}, for both TS and DCTS rewards decreased right after the change of user's preference. The performance of DCTS recovered around time step 250 while TS recovered around time step 300. This indicates that DCTS could learn the change of the user's preference more quickly than TS with the advantage of discounted rewards.

\subsection{Offline experiment using real-world data.}
In this section, we empirically compare DCTS with the TS, LinUCB with a hybrid linear model (hLinUCB) \cite{li2010contextual} and Transferable LinUCB (T-LinUCB) \cite{labille2021transferable} using real-world dataset. Details of baseline models are as follows:

\begin{itemize}
    \item hLinUCB \cite{li2010contextual}: LinUCB is an expansion of UCB policy and incorporates contextual features of arms. hybrid LinUCB incorporates contexts of both arms and users. 
    \item T-LinUCB \cite{labille2021transferable}: This model is an expansion of LinUCB. This calculates the prior evaluation of arms in the new domain with contextual feature of arms in different domains.
    \item Thompson Sampling \cite{chapelle2011empirical}: This model selects the action that maximizes the expected reward with respect to a randomly drawn belief of scores. In practice, this policy will sample scores from the posterior distribution and pull the arm with maximum score.
\end{itemize}

Because there was no public dataset that includes ads among multi-domains with contextual features of both users and ads, we leveraged our internal dataset at Rakuten. Rakuten is an e-commerce company in Japan that serves over 70 services, such as online shopping, financial services, and e-books. Among them, Rakuten Ichiba is an e-commerce platform and Rakuten Travel  is an online travel agency platform for lodging reservations. In this experiment, we used data from Rakuten Ichiba for pre-training and let DCTS learn the rewards of Rakuten Ichiba. We then evaluated performance for Rakuten Travel  and compared DCTS with other baselines. 

Regarding evaluation methods, there are various methods for offline policy evaluation of the bandit algorithm. Replay method \cite{li2011unbiased} is often leveraged but it requires random assignment of ads. Most datasets in the real world, however, cannot afford random assignment because of the counterfactual estimations. In other words, we can't know the true responses if we showed other ads to users. To overcome this issue, corrections such as Doubly Robust estimation \cite{jiang2016doubly} were proposed.  However, here we propose another simple way for the counterfactual estimation. In the real world, multiple ads are often displayed at the same time in the widget like a carousel. In this offline simulation, we regarded our model to pick only one of these ads to users at each step. The advantage of this method is: in this situation not only we know the user's response of that ad, but also we know user's responses for other ads because in the real-world these ads were displayed together. Moreover, for each impression of the carousel, we only have, at most, one clicked ad because otherwise users will see the landing page after clicking an ad. In this paper, we leveraged this technique and treated ads displayed at the same time in the carousel as displayed in the one slot to obtain counterfactual information.

\begin{table}
\caption{Statistics of Rakuten Ichiba and Rakuten Travel  data.}
\label{tab:statistic_data}
\centering
\begin{tabular}{ |c|c|c|c|c|}
\hline
&  
    impressions
 & 
    responses 
 & 
    \begin{tabular}{l}
    unique \\ users 
  \end{tabular}  
  & 
    ads
 \\
\hline
 \begin{tabular}{l}
    Rakuten \\ Ichiba
  \end{tabular} 
 & 191202 & 23394 & 33750 & 14\\
\hline
 \begin{tabular}{l}
    Rakuten \\ Travel
  \end{tabular}   & 16029 & 801 & 4552 & 5\\
\hline
\end{tabular}
\\
\end{table}

Utilizing the technique above, we showed one ad at each time step $\tau$, and if user $i$ clicked ad $k$ within 15 min, we regarded $s_{ik\tau}=1$ and $f_{ik\tau}=0$. Otherwise $s_{ik\tau}=0$ and $f_{ik\tau}=1$.  We discarded subsequent impressions within 5 min after the first impression from the same user to avoid the bias of repeated bot traffic. We summarized the Rakuten Ichiba data and Rakuten Travel  data we used in Table \ref{tab:statistic_data}. Please note that the sampled data does not represent the entire dataset. Datasets have a timestamp and we served them in chronological order to the model.

\begin{figure}
  \centering
  \begin{tabular}{c}
      \begin{minipage}[t]{1\hsize}
          \centering
          \includegraphics[width=7cm]{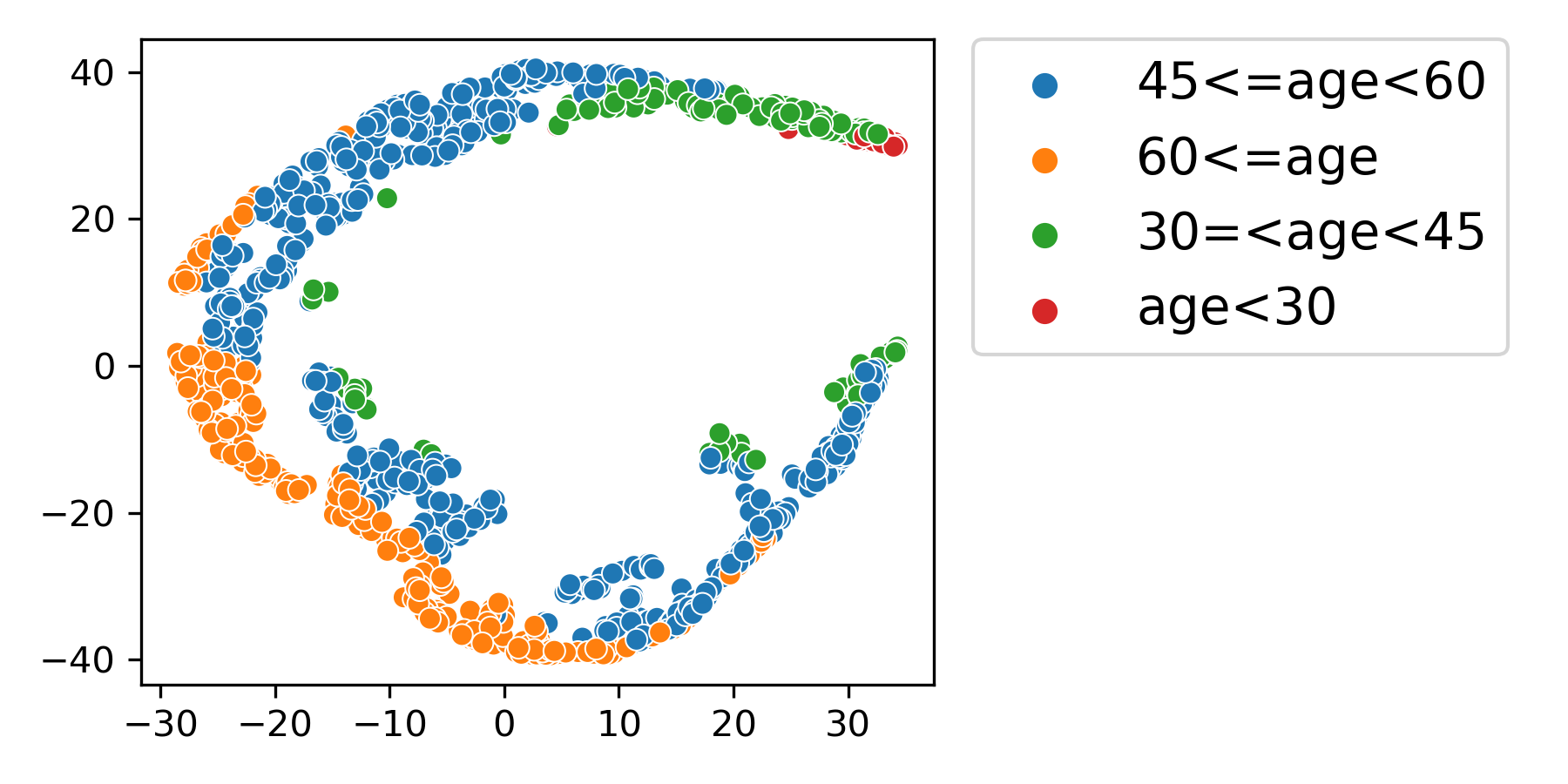}
          \caption{t-SNE projection of users labeled with age groups.}
          \small{Users in the same age group were clustered together.}

          \label{fig:sim_user}
      \end{minipage}

  \\
      \begin{minipage}[t]{1\hsize}
          \centering
          \includegraphics[width=7cm]{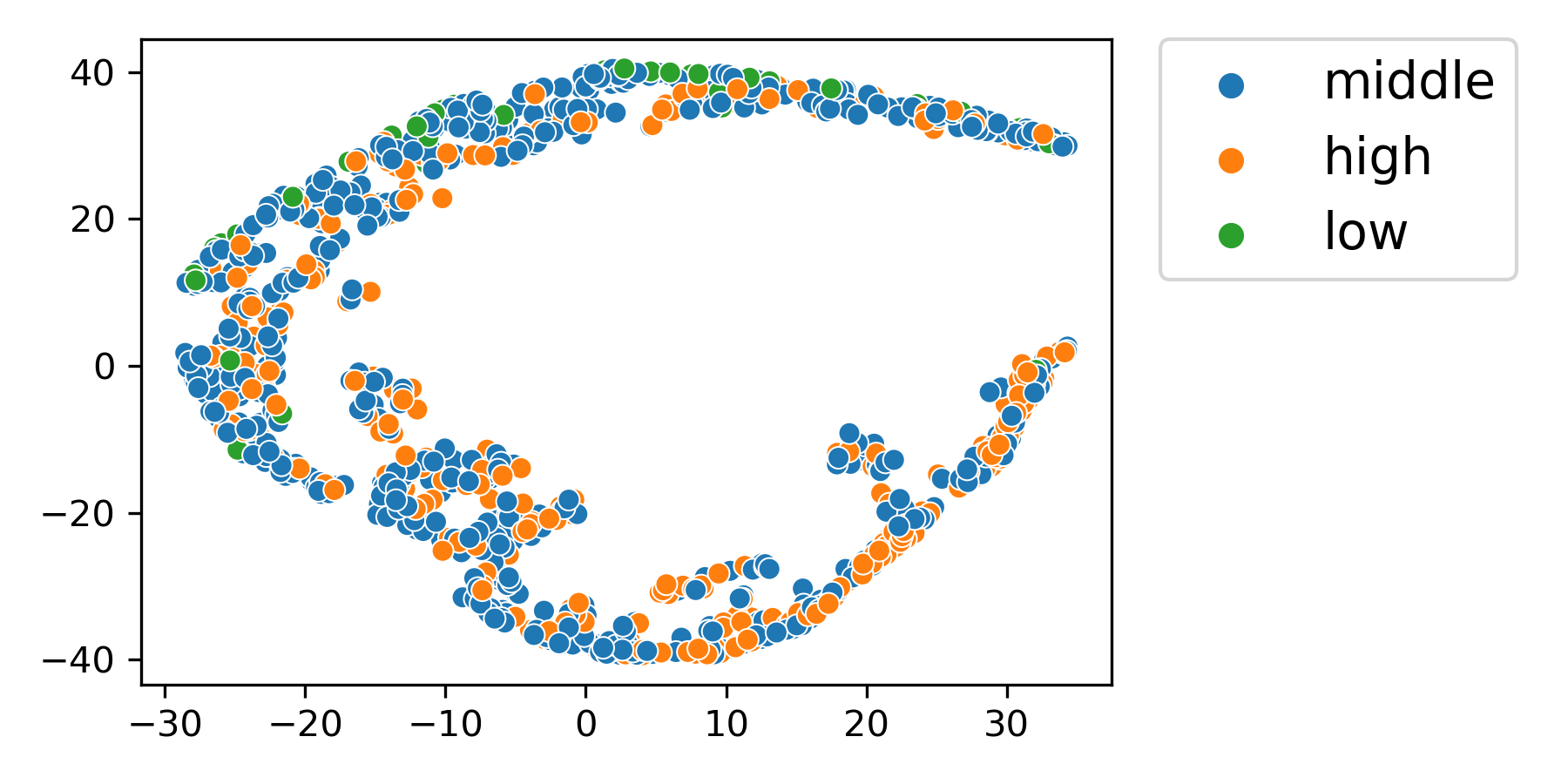}
          \caption{t-SNE projection of users labeled with reward point status.}
          \small{Users with low reward status were mostly placed in the upper part, while users with high reward status were placed in the lower part.}
          \label{fig:sim_point}
  
        \end{minipage}

   \\
      \begin{minipage}[t]{1\hsize}
         \centering
          \includegraphics[width=7cm]{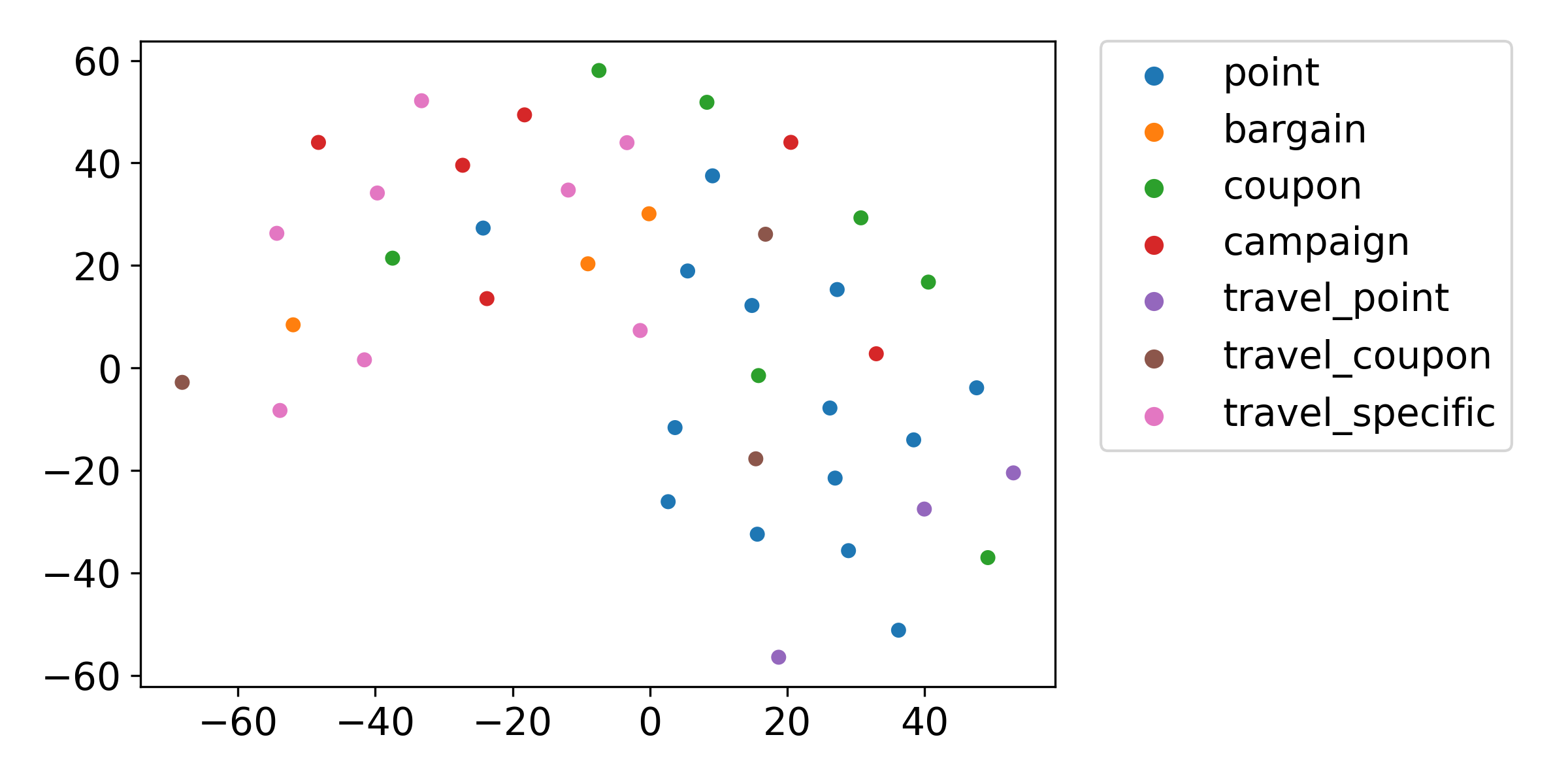}
          \caption{t-SNE projection of ads labeled with ad types.}
          \small{Labels without ``travel\_'' in the prefix means the ads in ``E-commerce service (Ichiba)''. Ads which are specific for travel such as promotions for sightseeing spots and hotels were clustered together. Also, point related ads for both Rakuten Ichiba and Travel were clustered together.}
             
          \label{fig:sim_ad}
  
        \end{minipage}

  \end{tabular}

\label{fig:tsne}
\end{figure}

DCTS also requires similarities among users and ads. To calculate them, we utilized Customer DNAs in Rakuten. Customer DNAs serves thousands of contextual attributes for each user. They consist of basic demographics such as gender and age, and users behaviors such as the preference of purchases and status of earned reward points. We leveraged 308 attributes among them and calculated the cosine similarity of users by Eq. \ref{eq:similarity}. We didn't have contextual attributes for ads so we regarded median values of contextual attributes of users who clicked the ads as the context of ads. We visualized the users and ads by t-SNE \cite{van2008visualizing} to check that the above similarities were reliable. As Fig. \ref{fig:sim_user} and Fig. \ref{fig:sim_point} showed, similarity of users could be largely explained by ages and point reward status. In other words, users with similar ages were placed closely. 
Also Fig. \ref{fig:sim_ad} showed similar type of ads were placed closely with each other in the latent structure. This implies that user and ad similarity indeed exist and can be utilized by DCTS.

\begin{figure}
  \centering
  \includegraphics[width=7cm]{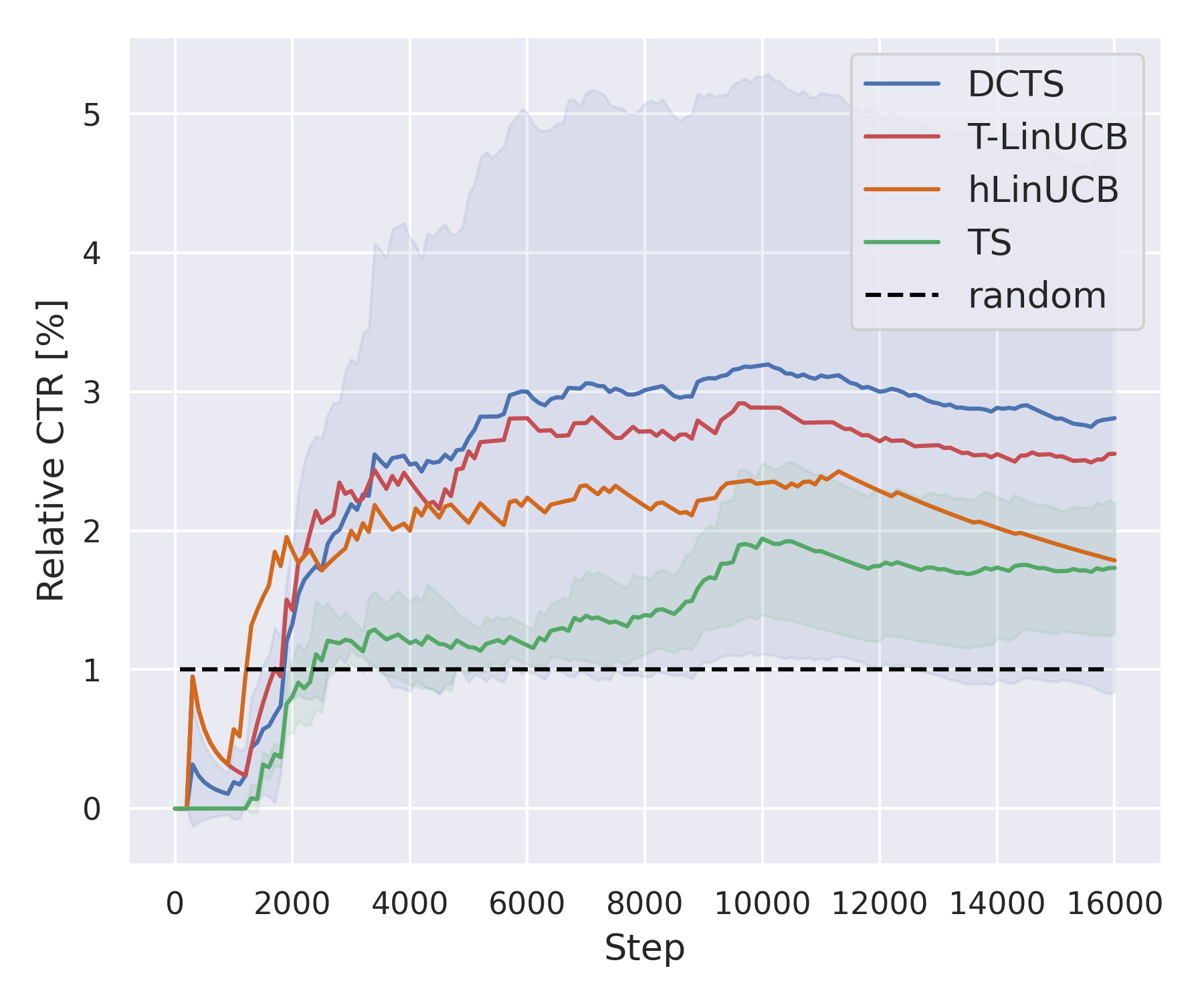}
  \caption{Relative CTR on Rakuten Travel  data. 
}
  \label{fig:industrial_result}
 \small{We compared dynamic collaborative filtering bandit (DCTS) with Transferable LinUCB (T-LinUCB), hybrid LinUCB (hLinUCB), Thompson Sampling (TS) and random. We initially pre-trained DCTS and T-LinUCB by Rakuten Ichiba data and then utilized them for Rakuten Travel data. }
\end{figure}

In Fig. \ref{fig:industrial_result}, we empirically verified that knowledge transfer among multi-domains, users, and temporal dynamics could improve the bandit model. Total CTRs of all policies were initially worse than random policy. Performance of hLinUCB improved at the step around 1000. At the step around 2000, T-LinUCB and DCTS outperformed hLinUCB and DCTS became the best, but DCTS had high variance. We observed improvement of performance at step around 9000 for TS and also observed that CTR of hLinUCB decreased after step around 11000. We currently assume this was because, at step 11000, customer preferences changed. As a result, DCTS performed better than T-LinUCB by 9.7 \% and better than LinUCB and TS by around 37 \%. We set DCTS's hyper-parameters as $\lambda=10$ and $g=1$ and $\gamma=0.5$ and discounted values by multiplying $\gamma$ every hour.

\begin{figure}
  \centering
  \includegraphics[width=7cm]{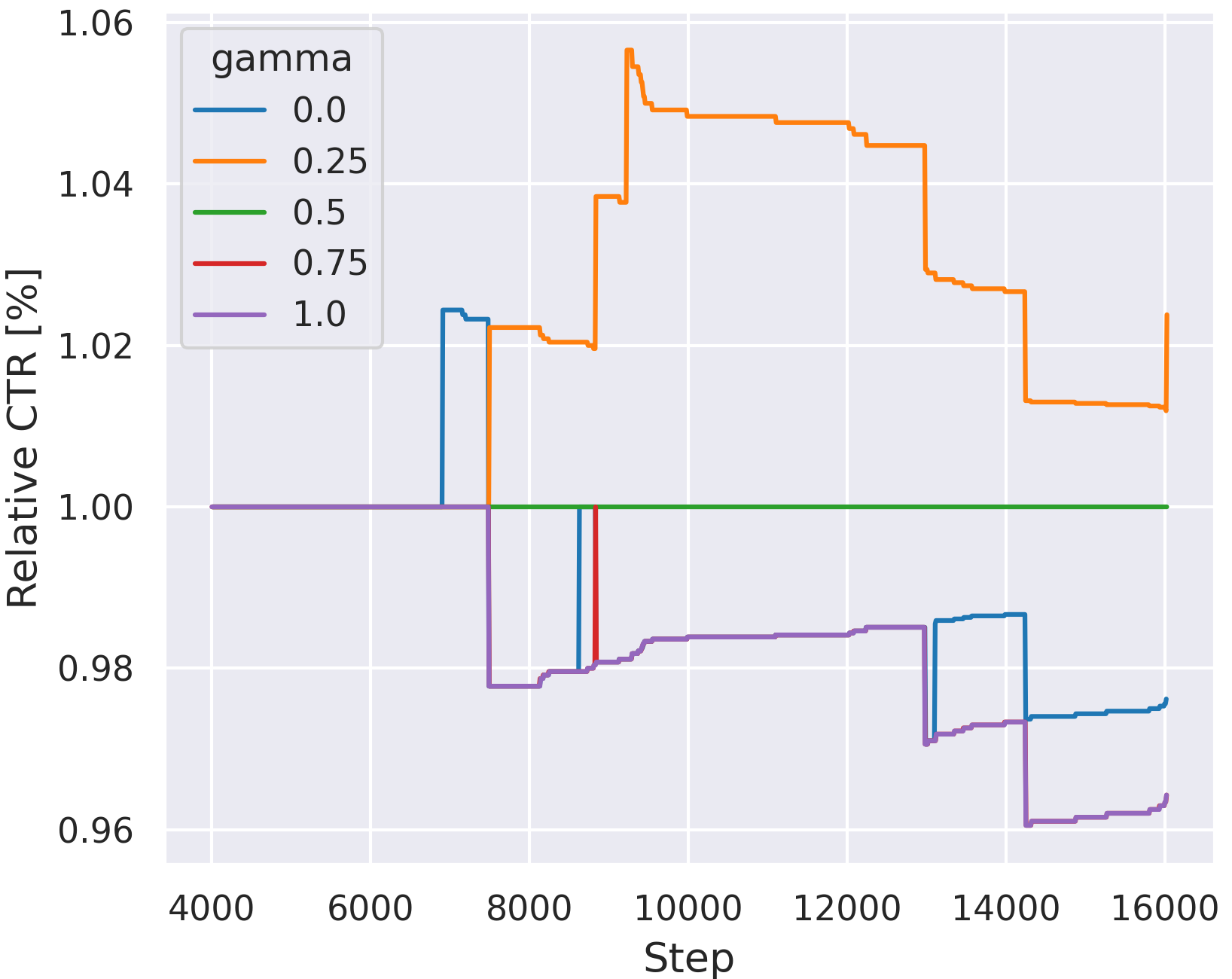}
  \caption{Relative CTR on Travel data with various $\gamma$ parameters}
  \label{fig:eval_decay}
\end{figure}

We also evaluated the performance change with different hyper-parameters. Fig. \ref{fig:eval_decay} shows relative CTRs by various $\gamma$, and we normalized values by the value of when $\gamma = 0.5$. According to Fig. \ref{fig:eval_decay}, we observed the best performance when $\gamma = 0.25$. It suggests this value is the most suitable decay for Rakuten Travel dataset. 

\section{Conclusion}

In this paper, we proposed a dynamic collaborative filtering Thompson sampling (DCTS) for a cross-domain and dynamic recommendation. DCTS leveraged the concept of discounting reward and collaborative filtering. As a result, it improved the prior distribution of Thompson sampling through leveraging similarity in users and ads. DCTS can adjust exploration and exploitation by choosing arm based on posterior probability and accelerate the convergence of the model, which led to maximizing cumulative rewards. We evaluated DCTS on the real-world data and showed that DCTS outperformed TS, hLinUCB, and random. We also plan to build a model which incorporates the pattern of the user's behaviour as a more accurate recommendation in the user's temporal dynamics.

\bibliographystyle{unsrt}
\bibliography{bibliography.bib}


\end{document}